\documentclass[preprint]{vldb}
\usepackage{graphicx}
\usepackage{balance}  
\usepackage{tcolorbox}

\newcommand{\remove}[1]{}

\usepackage{tikz}
\usetikzlibrary{shapes.misc}
\tikzset{cross/.style={cross out, draw=black, fill=none, minimum size=2*(#1-\pgflinewidth), inner sep=0pt, outer sep=0pt}, cross/.default={2pt}, extended line/.style={shorten >=-#1,shorten <=-#1}, extended line/.default=1cm, one end extended/.style={shorten >=-#1},
 one end extended/.default=1cm}
\usetikzlibrary{patterns}
\usetikzlibrary{positioning}
\usepackage[outline]{contour}
\contourlength{2pt}

\usepackage{url}
\usepackage{multirow}
\usepackage{enumitem}
\usepackage{array}

\vldbTitle{Piecewise Linear Approximation in Data Streaming Environments: Algorithmic Implementations and Experimental Analysis}
\vldbAuthors{Romaric Duvignau, Vincenzo Gulisano, Marina Papatriantafilou, Vladimir Savic}
\vldbDOI{https://doi.org/TBD}

\begin{document}



\title{Piecewise Linear Approximation in Data Streaming:\\ \Large\bf Algorithmic Implementations and Experimental Analysis \\ \Large}

\numberofauthors{2} 

\author{
%
%
\alignauthor
Romaric Duvignau, Vincenzo Gulisano, Marina Papatriantafilou \\
       \affaddr{Chalmers University of Technology}\\
       \email{\{duvignau,vinmas,ptrianta\}@chalmers.se}
\alignauthor
Vladimir Savic\titlenote{Work done while the author was affiliated with Chalmers University of Technology.}\\
       \affaddr{Qamcom R\&T}\\
       \affaddr{Gothenburg, Sweden}\\
       \email{vladimir.savic@qamcom.se}
}


\date{27 August 2018}

\toappear{}

\maketitle

\begin{abstract}
Piecewise Linear Approximation (PLA) is a well-established tool to reduce the size of the representation of time series by approximating the series by a sequence of line segments while keeping the error introduced by the approximation within some predetermined threshold. With the recent rise of edge computing, PLA algorithms find a complete new set of applications with the emphasis on reducing the volume of streamed data. In this study, we identify two scenarios set in a data-stream processing context: data reduction in sensor transmissions and datacenter storage. In connection to those scenarios, we identify several streaming metrics and propose streaming protocols as algorithmic implementations of several state of the art PLA techniques. In an experimental evaluation, we measure the quality of the reviewed methods and protocols and evaluate their performance against those streaming statistics. All known methods have deficiencies when it comes to handling streaming-like data, e.g. inflation of the input stream, high latency or poor average error. Our experimental results highlight the challenges raised when transferring those classical methods into the stream processing world and present alternative techniques to overcome them and balance the related trade-offs.
\end{abstract} 

\section{Introduction}\label{sec:introduction}

Computing a Piecewise Linear Approximation (PLA) of time series is a classical problem which asks to represent a series of timestamped records by a sequence of line segments while keeping the error of the approximation within some acceptable error bound. We consider in this study the online version of the problem with a prescribed maximum error~$\varepsilon$, \textit{i.e.} the time series is processed one record at a time, the output line segments are expected to be produced along the way, and the compressed line segments always fall within $\varepsilon$ from the original tuples. In the extensive literature dealing with such an approximation (among others \cite{elmeleegy2009online, hakimi1991fitting, luo2015piecewise, tomek1974two, xie2014maximum, zhao2016segmenting}), it is clearly stated that its main intend is to reduce the size of the input time series for both efficiency of storage and (later) processing. This entails a practical trade-off between a bounded precision loss and space saving for the time series representation. The advent of edge computing puts a brand new focus on this well-studied problem with new preoccupations, rather than merely only reducing the number of line segments of the PLA as in most previous works.

In this paper, we revisit this classical problem bearing in mind one of its main and most promising application: compression of numerical timestamped data streams. The present work, through thorough experimental analysis, shows that surprisingly, when confronted to reality, known algorithms tend to fail in various crucial aspects concerning their intended objective: they might inflate data, cause important delays in data propagation or result in higher than necessary error rates. In response to this, we also present alternatives that circumvent those drawbacks and reach better overall trade-offs, by carefully defining different protocols for producing compression records, as well as by introducing a new heuristic method to compute a PLA focusing on low errors. 

The recent constant increase in the quantity of continuous data produced by connected devices triggers numerous \textit{big data} challenges, and sets an environment where reducing data stream volume is more meaningful than ever. High speed production of data records induced by cyber-physical systems such as smart grids, homes and cars, moreover advocates fast \textit{online} algorithms. In this context, we identify two interesting scenarios for compression of streaming data.

\textbf{(1) Sensor transmission reduction:} In the first scenario, the primary goal of the compression is to reduce the communication between some small computing devices, that continuously sense data, and a data center collecting that information and monitoring a large swarm of such sensors. Here, instead of sending a new data tuple (i.e. a new data point) every time such a tuple is produced/sensed, only information concerning compressed segments of the PLA of the stream have to be sent once produced. By this way, we save the sensor's limited energy and we reduce the burden on the communication network by eliminating some communications between the two entities. The drawback is the introduction of a new \textbf{reconstruction delay} between the time the data is sensed and the time the data center can recover it from the sensor \textit{compressed stream}. Hence, one particular aspect to take into account while evaluating PLA techniques in streaming environments is this additional \textbf{latency} introduced by compressing the stream. In many practical scenarios such as anomaly detection, the data center needs to be able to reconstruct data relatively quickly after the data has been sensed. In our setting, the energy consumption due to increase in computation is considered negligible in regard to reduction in communications.

\textbf{(2) Datacenter storage reduction:} In the second scenario, a data center receives a large quantity of streamed data, too massive to be processed on-line and decides to store some of its data streams upon their reception. In this scenario, compression algorithms should have small time overhead per tuple to deal with the potentially high speed of the input streams, and should aim to optimize the total \textbf{storage space} used to represent compressed data. 

These scenarios form a common framework where an input stream is processed and a compressed stream is generated and transmitted or stored, to be later (or on reception) reconstructed in an online fashion. In both scenarios, the use of PLA to \textit{compress} the input data stream is tied to a potential compromise between the quantity of \textbf{precision lost} on the input stream and the achieved qualitative compression: a better compression entails fewer transmissions in our first scenario, and a smaller data volume in our second scenario. Our experimental setting will show, however, that reaching the best compression rate should not always be seen as a golden target for PLA algorithms. In particular, by playing with the actual PLA method used to produce the compression stream, we will see that a relatively small loss on the compression rate may entail great improvements on the average precision of the reconstructed stream.

It is important for us to highlight that in our context and in particular our first use case, decision making should be considered irreversible, as once the data (e.g. information about a new line segment) has been transmitted over the network, it is hardly modifiable. Sending more data to alter previous decisions may significantly perturb the processing of data on the server side of the transmission, and is therefore often not applicable.

These two scenarios lead us to investigate the current state of the art solutions for constructing PLA of timestamped data streams with respect to the following three aspects:
\begin{itemize}[topsep=5pt,itemsep=0pt,parsep=5pt]
	\item \textbf{Compression:} capturing the space gain between the compressed stream and the original input stream;
	\item \textbf{Latency:} capturing the delay in number of records introduced by the compression process (cf \S~\ref{subsec:performance_metrics});  
	\item \textbf{Error:} capturing the precision loss by considering continuous and average error rather than only maximum error as in previous works (see Section~\ref{sec:relatedwork}).
\end{itemize}

In this paper, we evaluate four PLA techniques representing the most common ones found in the literature with respect to these three criteria. The evaluation is conducted against datasets of recent real-world streamed data (from different sensor types such as GPS, lasers, or speedometer) as well as a dataset used in several previous works on PLA. Our experimental results show relatively poor trade-offs in regard to our streaming metrics for historical methods.

We further investigate two of these methods that fit a streaming-oriented PLA processing mechanism introduced with this survey, as well as a new heuristic one. The results demonstrate significant improvements in the three sought aspects compared to the aforementioned methods. An important contribution of this paper is to clearly define a complete framework for measuring performance of PLA compression in streaming environments, which may ease the development of new advances and research in this domain.

A need for simple, efficient and adaptive solutions is implied in both scenarios. This is particularly important in our first scenario where the device-limited computing power calls for simple algorithmic solutions with small memory requirement. In our second scenario, a small computing overhead to calculate the compressed stream is naturally sought in contemporary data centers. However, note that in this present survey we do not deal with the computing overhead introduced by the compression process.

The rest of the paper is organized as follows. The next section is dedicated to introducing precise definitions for PLA of time series as used in this paper. In Section~\ref{sec:plamethods}, four classical PLA techniques are reviewed (named here \textit{Angle}, \textit{Disjoint}, \textit{Continuous} and \textit{MixedPLA}), together with a new one (\textit{Linear}) introduced in this article. We formalize the notion of a streaming PLA compression algorithm and identify different streaming metrics to evaluate them in Section~\ref{sec:streamingpla}. Section~\ref{sec:protocols} introduces different algorithmic implementations in order to generate in an online fashion the compression records that will be either stored or transmitted. The subsequent section presents our core results, a thorough evaluation of the existing and new methods against real world traces for different kinds of timestamped data streams. In Section~\ref{sec:relatedwork}, related work in the literature that leads to this paper is reviewed. The last section presents our conclusions and different future directions for the work initialized by this evaluation study.

\section{Preliminaries}\label{sec:preliminaries}
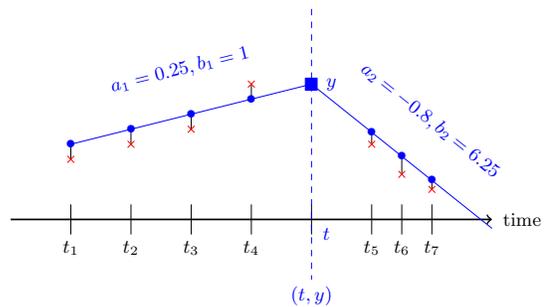
\begin{figure}[t]
	\centering
	\scalebox{0.8}{%
    \begin{tikzpicture}
    	\draw[->, thick] (0,0) -- (8,0) node[xshift=0.5cm] {time};
    	\draw (1,0.25) -- (1,-0.25) node[below] {$t_1$};
    	\draw (2,0.25) -- (2,-0.25) node[below] {$t_2$};
    	\draw (3,0.25) -- (3,-0.25) node[below] {$t_3$};
    	\draw (4,0.25) -- (4,-0.25) node[below] {$t_4$};
    	\draw (6,0.25) -- (6,-0.25) node[below] {$t_5$};
    	\draw (6.5,0.25) -- (6.5,-0.25) node[below] {$t_6$};
    	\draw (7,0.25) -- (7,-0.25) node[below] {$t_7$};
    	
    	\draw[blue] (1,1.25) -- (5,2.25) node[xshift=0.33cm,blue] {$y$} node[pos=0.5,sloped,yshift=0.5cm,above] {$a_1=0.25,b_1=1$};
    	
    	\draw (1,1) -- (1,1.25);
    	\draw (2,1.25) -- (2,1.5);
    	\draw (3,1.5) -- (3,1.75);
    	\draw (4,2.25) -- (4,2);
    	\draw (6,1.25) -- (6,1.45);
    	\draw (6.5,0.75) -- (6.5,1.05);
    	\draw (7,0.5) -- (7,0.65);
    	
    	\node[cross, red] at (1,1) {};
    	\node[cross, red] at (2,1.25) {};
    	\node[cross, red] at (3,1.5) {};
    	\node[cross, red] at (4,2.25) {};
    	\node[cross, red] at (6,1.25) {};
    	\node[cross, red] at (6.5,0.75) {};
    	\node[cross, red] at (7,0.5) {};
    	
    	\node[blue] at (1,1.25) {$\bullet$};
    	\node[blue] at (2,1.5) {$\bullet$};
    	\node[blue] at (3,1.75) {$\bullet$};
    	\node[blue] at (4,2) {$\bullet$};
    	
    	\node[blue] at (5,2.25) {$\blacksquare$};

    	\draw[blue] (5,2.25) -- (8,-0.15) node[pos=0.5,sloped,yshift=0.5cm,above] {$a_2=-0.8,b_2=6.25$};
    	
    	\node[blue] at (6,1.45) {$\bullet$};
    	\node[blue] at (6.5,1.05) {$\bullet$};
    	\node[blue] at (7,0.65) {$\bullet$};
    	
    	\draw (5,0.25) -- (5,-0.25);
    	\node[blue] at (5.25,-0.25) {$t$};
    	
    	\draw[blue, dashed] (5,3.5) -- (5,-1) node[below] {$(t,y)$};
    \end{tikzpicture}
    }
    \caption{A joint knot $(t,y)$ PLA record.}
	\label{fig:jointknot}
\end{figure}

In this section, we describe the problem of computing a piecewise linear approximation of a data stream while tolerating a predetermined maximum error of $\varepsilon$. We adopted and formalized here the nomenclature introduced in~\cite{luo2015piecewise} such that all PLA methods evaluated in this work fit a common framework. A (potentially infinite) stream $(t_i,y_i)_{i \geq 0}$ of points is read as input one tuple at a time, and along the processing of inputs a series of \textit{PLA records} $(r_j)_{j \geq 0}$ is generated. We assume $t_i$'s representing \emph{timestamps} are strictly increasing, while $y_i$'s can vary arbitrarily. To not burden the notation, we denote  the sequences of tuples as $(t_i,y_i)_{i \geq 0}$ in place of the perhaps more conventional $((t_i,y_i))_{i \geq 0}$ notation.

\begin{figure}
    \centering
    \scalebox{0.8}{%
    \begin{tikzpicture}
    	\draw[->, thick] (0,0) -- (8,0) node[xshift=0.5cm] {time};
    	\draw (1,0.25) -- (1,-0.25) node[below] {$t_1$};
    	\draw (2,0.25) -- (2,-0.25) node[below] {$t_2$};
    	\draw (3,0.25) -- (3,-0.25) node[below] {$t_3$};
    	\draw (4,0.25) -- (4,-0.25) node[below] {$t_4$};
    	\draw (6,0.25) -- (6,-0.25) node[below] {$t_5$};
    	\draw (6.5,0.25) -- (6.5,-0.25) node[below] {$t_6$};
    	\draw (7,0.25) -- (7,-0.25) node[below] {$t_7$};
    	
    	\draw[blue] (1,1.25) -- (5,3.25) node[xshift=0.33cm,blue] {$y'$} node[pos=0.5,sloped,yshift=0.5cm,above] {$a_1=0.5,b_1=0.75$};
    	
    	\draw (1,1) -- (1,1.25);
    	\draw (2,2) -- (2,1.75);
    	\draw (3,2.5) -- (3,2.25);
    	\draw (4,3) -- (4,2.75);
    	\draw (6,1.25) -- (6,1);
    	\draw (6.5,0.75) -- (6.5,0.9);
    	\draw (7,0.5) -- (7,0.8);
    	
    	\node[cross, red] at (1,1) {};
    	\node[cross, red] at (2,2) {};
    	\node[cross, red] at (3,2.5) {};
    	\node[cross, red] at (4,3) {};
    	\node[cross, red] at (6,1.25) {};
    	\node[cross, red] at (6.5,0.75) {};
    	\node[cross, red] at (7,0.5) {};
    	
    	\node[blue] at (1,1.25) {$\bullet$};
    	\node[blue] at (2,1.75) {$\bullet$};
    	\node[blue] at (3,2.25) {$\bullet$};
    	\node[blue] at (4,2.75) {$\bullet$};
    	
    	\node[blue] at (5,3.25) {$\blacksquare$};
    	
    	\draw[blue] (5,1.2) node[xshift=-0.33cm,blue] {$y''$} -- (8,0.6) node[pos=0.5,sloped,yshift=0.5cm,above] {$a_2=-0.2,b_2=2.2$};

    	\node[blue] at (5,1.2) {$\blacksquare$};
    	
    	\node[blue] at (6,1) {$\bullet$};
    	\node[blue] at (6.5,0.9) {$\bullet$};
    	\node[blue] at (7,0.8) {$\bullet$};
    	
    	\draw (5,0.25) -- (5,-0.25);
    	\node[blue] at (5.25,-0.25) {$t$};
    	
    	\draw[blue, dashed] (5,3.5) -- (5,-1) node[below] {$(t,y',y'')$};
    \end{tikzpicture}
    }
    \caption{A disjoint knot $(t,y',y'')$ PLA record.}
	\label{fig:disjointknot}
\end{figure}
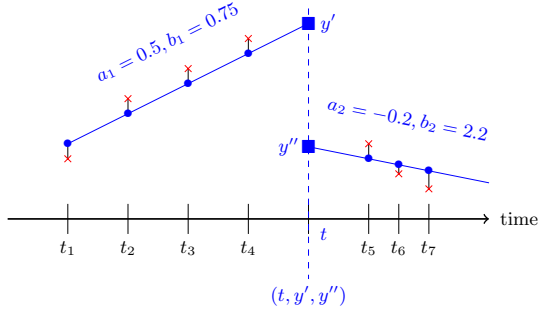

PLA records $(r_j)_{j \geq 0}$ (Figures~\ref{fig:jointknot} and~\ref{fig:disjointknot}) encode the input stream's segment line approximation and can be of: 
\begin{itemize}
    \item A \textbf{joint knot} form, consisting of a pair of values $(t,y)$; such a record encodes two consecutive line segments sharing the same endpoint $(t,y)$, that is both the ending point of the first segment and the starting point of the following one.
    \item A \textbf{disjoint knot} form, consisting of a triplet of values $(t,y',y'')$; such a record encodes two consecutive segments with different endpoints sharing the same $t$-value -- the first one ends in $(t,y')$ while the next one starts in $(t,y'')$.
\end{itemize}

The \textbf{sequence of PLA records} $(r_j)_{j \geq 0}$ must be strictly increasing in its $t$-values (first field of both record types), and the first knot of the sequence and the last knot (if the sequence is finite) are both joint knots. Moreover, for each PLA record $r_j$ for $j > 0$, we can associate a line segment $(a_j, b_j)$ in the following manner. Let us define first the two endpoints $(u_j,v_j)$ and $(u'_j,v'_j)$ of the $j$-th line segment as
\begin{align*}
  (u_j,v_j)&=\begin{cases}
    (t,y) & \text{if $r_{j-1}$ is a joint knot $(t,y)$},\\
    (t,y'') & \text{if $r_{j-1}$ is a disjoint knot $(t,y',y'')$},
  \end{cases} \\
  (u'_j,v'_j)&=\begin{cases}
    (t,y) & \text{if $r_{j}$ is a joint knot $(t,y)$},\\
    (t,y') & \text{if $r_{j}$ is a disjoint knot $(t,y',y'')$}.
  \end{cases}
\end{align*}

Then $(a_j, b_j)$ are respectively the slope and $y$-intercept of the line segment connecting $(u_j,v_j)$ and $(u'_j,v'_j)$, \textit{i.e.} $$a_j = \frac{v'_j-v_j}{u'_j-u_j} \quad\text{ and }\quad b_j = \frac{u'_j v_j - v'_j u_j}{u'_j-u_j}.$$

The sequence of coefficients $(a_j, b_j)_{j > 0}$ associated with a PLA record stream allows us to \emph{reconstruct} an approximate version of the input stream from the original timestamps $(t_i)_{i \geq 0}$. We define the $i$-th reconstructed record $(t'_i, y'_i)$ as
$$(t_i, a_j t_i + b_j) \quad\text{ with $j = \max\limits_{\ell > 0} \{ \ell \;|\; t_i \geq u_\ell \}$ }.$$

Said otherwise, the input timestamps are processed in order and for each one, a reconstructed tuple can be computed based on which approximation line segment the original timestamp falls within. For each PLA record $r_j$, we can thus associate a finite sequence $\hbox{reconstruct}(r_j)$ of $t$-consecutive tuples from the reconstructed stream: those values are exactly the ones reconstructed from the $j$-th line segment. 

We call \textit{error} $e_i$ at input point $(t_i, y_i)$ the absolute difference between the approximated point $(t'_i, y'_i)$ and its real counterpart, \textit{i.e.} $e_i=|y_i-y'_i|$. Such errors are commonly referred as \emph{residuals} in statistics terminology. Finally, a \textit{PLA method} for error threshold $\varepsilon$ is an algorithm that reads as input a stream of $(t_i,y_i)_{i \geq 0}$ and outputs a stream of PLA records $(r_j)_{j \geq 0}$, such that $e_i < \varepsilon$ for all $i \geq 0$. 

Figures~\ref{fig:jointknot} and \ref{fig:disjointknot} present an example of PLA where original values from the input stream are marked as crossed, reconstructed tuples are bulleted, PLA knots are squares, errors due to compression are shown as vertical bars between both tuples, and linear coefficients $a$ and $b$ are displayed over approximation line segments.

Once this setting is set, the goal of PLA methods developed in the literature for decades has been to minimize the number of output line segments for a given maximum error $\varepsilon$ on a finite stream of input tuples, \textit{i.e.} to minimize the length of the sequence $r_j$ of PLA records. Other methods have been devised to enhance a better throughput or optimize some other measure of fitness (cf next section for a description of the main ones found in the recent literature). 

In the following sections, we often explain that we process \emph{error segments} for a particular error threshold $\varepsilon$; those are simply the 2D closed line segments formed by the two extreme approximated values allowed for a particular input point $(t,y)$, \textit{i.e.} the error segment associated with tuple $(t,y)$ is the line segment $\overline{(t, y-\varepsilon),(t, y+\varepsilon)}$, where $(t, y-\varepsilon)$ will be usually referred as its lower endpoint and $(t, y+\varepsilon)$ as its upper endpoint. In order to form a correct piecewise linear approximation, PLA segments produced by some algorithm have to intersect every single error segment (such that the reconstructed value for that particular timestamp is within the tolerated limits).

\section{PLA methods}\label{sec:plamethods}

\begin{table*}[t!]
    \centering
    \begin{tabular}{c|c|c|c|c|c|c|c}
        \textbf{PLA method} & \textbf{Complexity} & \textbf{Latency} & \textbf{Proc. time} & \textbf{Opt. criteria} & \textbf{Size of rec.} & \textbf{No. segments} & \textbf{Compr.}\\
        \hline
        SwingFilter \cite{elmeleegy2009online} & $\mathcal{O}(1)$ & $++$ & $++$ & none & $2$ & $--$ & $--$ \\
        Angle \cite{xie2014maximum} & $\mathcal{O}(1)$ & $+$ & $++$ & none & $3$ & $-$ & $-$ \\
        
        Disjoint \cite{elmeleegy2009online,xie2014maximum} & $\mathcal{O}(n) / \mathcal{O}(1)$ & $+/-$ & $+/-$ & disj. knots & $3$ & $+$ & $+$ \\
        
        Continuous \cite{imai1986optimal,hakimi1991fitting} & $\mathcal{O}(n) / \mathcal{O}(1)$ & $-$ & $+/-$ & joint knots & $2$ & $-$ & $+$ \\
        
        MixedPLA \cite{luo2015piecewise} & $\mathcal{O}(n) / \mathcal{O}(1)$ & $--$ & $--$ & global size & $2$ and $3$ & $++$ & $++$ \\
        
        Linear (new) & $\mathcal{O}(n) / \mathcal{O}(n)$ & $+$ & $+$ & none & $3$ & $+/-$ & $+/-$
    \end{tabular}
    \caption{Summary of the evaluated PLA methods and their respective features: worst case/amortized complexity, latency (as defined in \S~\ref{subsec:performance_metrics}), processing time, optimality criteria, size of records, number of segments and overall compression on a relative scale $--$ (worst), $-$, $+/-$, $+$, $++$ (best).}
    \label{tab:plamethods}
\end{table*}

We shall have in this section a brief overview of each evaluated PLA method; their characteristics are summarized in Table~\ref{tab:plamethods}. For a historical review of the methods, please refer to Section~\ref{sec:relatedwork} and for a complete description of each method, to the references therein. The name of each PLA method in the table is the one used later in our evaluation in \S~\ref{sec:evaluation}. Note that this summary table has been filled based partially on results from preceding experimental evaluations~\cite{elmeleegy2009online,xie2014maximum,luo2015piecewise}, complemented by parts of our own results. The methods outputting only disjoint knots (\textit{i.e.} angle, disjoint, and linear) will be able to fit in a stream-oriented output framework introduced in Section~\ref{sec:protocols}.

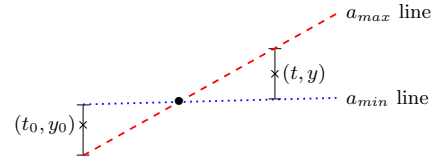
\begin{figure}[b]\centering
    \scalebox{0.85}{%
    \begin{tikzpicture}[yscale=0.8]
    \node[cross] (0) at (0,0) {};
    \draw (0,0) node[left] {$(t_0,y_0)$};
    \draw[|-|] (0,0.4) -- (0,-0.6);
    
    \node[cross] (1) at (3,1) {};
    \draw (3,1) node[right] {$(t,y)$};
    \draw[|-|] (3,1.5) -- (3,0.5);
    
    \draw [thick, dashed, red] (0,-0.6) -- (4,2.2);
    \draw [thick, dotted, blue] (0,0.4) -- (4,0.53);
    
    \draw (4,2.2) node[right] {$a_{max}$ line};
    \draw (4,0.53) node[right] {$a_{min}$ line};
    
    \draw (1.5,0.45) node {$\bullet$};
    \end{tikzpicture}
    }
    \caption{Bounding lines: the minimum-slope is represented dotted whereas maximum-slope is dashed.}\label{fig:slopes}
\end{figure}

Until recently, PLA methods (presented in \S~\ref{greedy} -- \S~\ref{continuous}) have been following a common approach: a longest (for some criteria) possible approximation line is constructed from an initial point, till encountering a \textbf{break-up point} $(t,y)$ that cannot be approximated with the previous sequence of points. The new initial point then is the break-up point  (creating disjoint knots) or the end part of the previous segment (for joint knots).

This may be referred to as ``greedy approximation procedure'', since each decision is instantly taken when reviewing the input point and is irreversible. In a different approach,  given in~\cite{luo2015piecewise}, the decision is not taken when a point is processed but later on, when the program can decide that the optimal approximation for a given fitness function cannot be modified by the addition of new input points. The latter is described in~\S~\ref{mixedPLA}.

All PLA methods revolve around calculating extreme slope lines in order to quickly evaluate if a new input point might break or not the current approximation (an example is depicted on Figure~\ref{fig:slopes}). Indeed, those extreme lines representing the maximum and minimum possible slopes for the current approximation, bound the possibilities for adding new data points: if the new additional tuple lies outside of the zone delimited by the extreme lines, the current approximation cannot be pursued.
Different methods, focusing on different optimization goals, have been devised to efficiently and accurately recompute them after each input arrival. 

\subsection{Angle-based Greedy Approximations} \label{greedy}
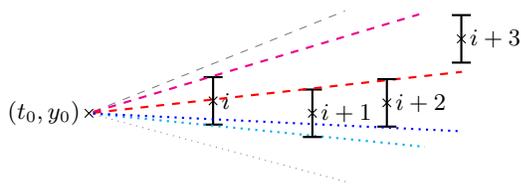
\begin{figure}[b]\centering
\begin{tikzpicture}[scale=0.33]
\node[cross] (0) at (0,0) {};
\draw (0,0) node[left] {$(t_0,y_0)$};

\node[cross] (i) at (5,0.5) {};
\draw (5,0.5) node[right] {$i$};
\draw[|-|, thick] (5,1.5) -- (5,-0.5);

\draw [dashed, gray, one end extended=1cm] (0) -- (7.5,3);
\draw [dotted, gray, one end extended=1cm] (0) -- (7.5,-2);

\draw [magenta, dashed, thick, one end extended=2cm] (0) -- (7.5,2.25);
\draw [cyan, dotted, thick, one end extended=2cm] (0) -- (7.5,-0.75);

\node[cross] (1) at (9, 0) {};
\draw (9,0) node[right] {$i+1$}; 
\draw[|-|, thick] (9,1) -- (9,-1);

\draw [red, dashed, thick, one end extended=2cm] (0) -- (9,1);

\node[cross] (2) at (12, 0.42) {};
\draw (12,0.42) node[right] {$i+2$}; 
\draw[|-|, thick] (12,1.42) -- (12,-0.58);

\draw [blue, dotted, thick, one end extended=1cm] (0) -- (12,-0.58);

\node[cross] (3) at (15, 3) {};
\draw (15,3) node[right] {$i+3$}; 
\draw[|-|, thick] (15,4) -- (15,2);

\end{tikzpicture}
\caption{Illustration of the angle PLA technique; here $i+3$ is the \textit{break-up} point.}\label{fig:swingfilter}
\end{figure}

The simplest method, \textit{SwingFilter}~\cite{elmeleegy2009online}, simplifies the problem of calculating a compressed line segment by fixing beforehand the origin of the compressed segment; once the origin is set, it becomes straightforward to construct the longest possible line segment within the error threshold in a constant number of operations per input point. 

The method relies on a simple ``angle'' or ``swing'' computation illustrated on Figure~\ref{fig:swingfilter}. Suppose one wants to construct an approximation line segment starting from $(t_0,y_0)$. The first angle formed by the bounding slope lines is trivially computed from $(t_0,y_0)$, the following tuple $(t_1,y_1)$ and $\varepsilon$. Any incoming tuple $(t,y)$ modifies the current validity angle by increasing/reducing one or both bounding lines if $(t,y)$ falls at least partially within the angle, and terminates the approximation otherwise. When such a break-up point is encountered, the compression ends and a line segment is output\footnote{From the method's point of view, the choice of the line to output at that point does not matter, but may influence other metrics as approximation errors.}. The compression restarts then with the endpoint of the previous segment as new origin, which generates a PLA of joint knots. 

In this study, we also use  a variant of this technique introduced in~\cite{xie2014maximum} and named here \textit{Angle}: instead of initializing the origin of the angle to the first point to approximate, the intersection of the maximum and minimum sloping line is taken ($\bullet$ symbol in Figure~\ref{fig:slopes}). This point has the advantage of offering a wider angle than the first point to compress $(t_0,y_0)$ while still assuring that any chosen slope within the angle will intersect with $(t_0,y_0)$'s error segment. Also, the compression is restarted from the break-up point, hence generating a stream of disjoint knots.

The method is important not only for its simplicity but also for its \textit{worst-case} $\mathcal{O}(1)$ processing time per node, which can be a significant advantage in many applications; this is however to be contrasted by it presenting the poorest compression.

\subsection{Optimal Disjoint Approximation} \label{disjoint}
\begin{figure}[b]\centering
    \begin{tikzpicture}[scale=0.33]
    \node[cross] (1) at (0,0) {};
    \node[cross] (2) at (1,1) {};
    \node[cross] (3) at (3,2) {};
    \node[cross] (4) at (6,3) {};
    \node[cross] (5) at (9,3.5) {};
    \node[cross] (6) at (12,2) {};
    \node[cross] (7) at (15,-1) {};
    \node[cross] (8) at (10,-3) {};
    \node[cross] (9) at (5,-2) {};
    \node[cross] (10) at (2,-1) {};
    \draw[cyan, dashed, thick] (1)--(2)--(3)--(4)--(5);
    \draw[blue, dashed, thick] (5)--(6)--(7);
    \draw[red, dotted, thick] (7)--(8);
    \draw[magenta, dotted, thick] (8)--(9)--(10)--(1);
    \node[cross] at (0.75,0.5) {};
    \node[cross] at (0.33,0.1) {};
    \node[cross] at (1.5,1) {};
    \node[cross] at (2.5,-0.5) {};
    \node[cross] at (4,0.33) {};
    \node[cross] at (4.5,-1) {};
    \node[cross] at (5.5,2) {};
    \node[cross] at (6.5,0) {};
    \node[cross] at (7.5,-2) {};
    \node[cross] at (9.33,0) {};
    \node[cross] at (8.5,2) {};
    \node[cross] at (9.5,2) {};
    \node[cross] at (10.5,-1) {};
    \node[cross] at (11.5,1) {};
    \node[cross] at (12.5,0) {};
    \node[cross] at (13.5,-1) {};
    \node[cross] at (14.5,-1) {};
    \end{tikzpicture}
    \caption{Example of a convex hull.}\label{fig:hull}
\end{figure}
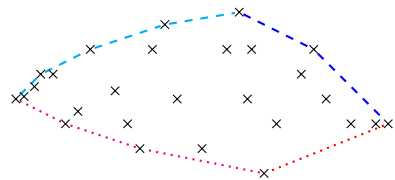

A second approach aims at finding the optimal PLA, in number of line segments, using only disjoint knots. The problem corresponds to finding the longest possible approximation segment, till a break-up point is found, then restart the algorithm from that tuple (see e.g.~\cite{xie2014maximum}).  It is therefore similar in nature to the greedy approximation without the constraint on fixing \textit{a priori} an origin for the valid angle.

From a fixed starting point, the longest approximation segment can be calculated in $\mathcal{O}(1)$ amortized time per point, by maintaining two convex hulls while processing new inputs. Recall the \textit{convex hull} of a set of points is the minimum convex set containing all of them. It is further divided into two components: the \textit{upper hull} is the set of points above (\text{i.e.} with higher $y$-coordinate) the line between the leftmost and rightmost points (regarding their $t$-coordinate) of the convex hull, and the \textit{lower hull} the remaining points of the hull's envelope (Figure~\ref{fig:hull} gives an example where the upper hull is dashed, and lower one dotted).

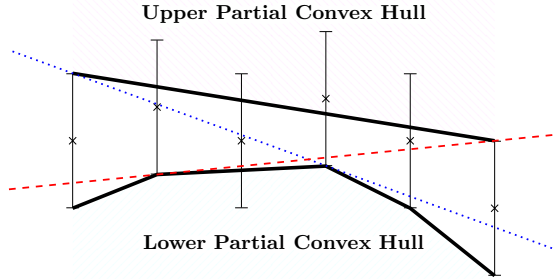
\begin{figure}[t]
    \centering
    \scalebox{0.85}{%
    \begin{tikzpicture}[scale=0.66, yscale=0.8]
    
    \node[circle,transparent] (f1) at (0,6.25) {};
    \node[circle,transparent] (f2) at (10,6.25) {};
    \node[circle,transparent] (f3) at (0,-3) {};
    \node[circle,transparent] (f4) at (10,-3) {};
    
    \fill[pattern=north west lines, pattern color=magenta, opacity=0.2] (0,4) to (f1.center) to  (f2.center) to (10,2) to cycle;
    
    
    \fill[pattern=north east lines, pattern color=cyan, opacity=0.2] (10,-2) to (f4.center) to (f3.center) to (0,0) to (2,1) to (6,1.25) to (8,0) to cycle;
    
    \node[cross] (0) at (0,2) {};
    \draw[|-|] (0,0) -- (0,4);
    
    \node[cross] (1) at (2,3) {};
    \draw[|-|] (2,1) -- (2,5);
    
    \node[cross] (2) at (4,2) {};
    \draw[|-|] (4,0) -- (4,4);
    
    \node[cross] (3) at (6,3.25) {};
    \draw[|-|] (6,1.25) -- (6,5.25);
    
    \node[cross] (4) at (8,2) {};
    \draw[|-|] (8,0) -- (8,4);
    
    \node[cross] (5) at (10,0) {};
    \draw[|-|] (10,-2) -- (10,2);
    

    
    \draw[ultra thick] (0,0) -- (2,1) -- (6,1.25) -- (8,0) -- (10,-2);
    \draw[ultra thick] (0,4) -- (10,2);
    
    \draw[dotted, blue, extended line, thick] (0,4) -- (10,-0.565);
    \draw[dashed, red, extended line, thick] (0,0.75) -- (10,2);
    

    \node[font=\bf] at (5,-1) {\contour{white}{Lower Partial Convex Hull}};
    \node[font=\bf] at (5,5.75) {\contour{white}{Upper Partial Convex Hull}};
    
    \end{tikzpicture}
    }
    \caption{Upper and lower partial convex hulls.}\label{fig:hulls}
    
\end{figure}

In order to dynamically maintain both bounding lines, one has to keep track of two particular convex hulls: the one formed by the lower endpoints of error segments (called \textit{lower convex hull}) and the one formed by their upper end points (\textit{upper convex hull}). More precisely, we are only interested in the ``outer enveloppe'' of those two hulls, and by that we mean the portion of the hull facing ``inwards'' the input points. Figure~\ref{fig:hulls} gives an example where the non-hatched zone delimits where approximation segments can be found. In the worst case scenario, the hulls might be as large as the number of processed input points, but in practice they are way smaller (cf \cite{xie2014maximum,luo2015piecewise}). By using a variation of the \textit{angle} technique, it is possible to update the hulls in constant amortized time per processed input \cite{elmeleegy2009online,xie2014maximum}. Once a break-up point is found, a segment line is generated\footnote{The method itself again does not enforce how this part is solved, but all implementations pick the ``average'' of the extreme slope lines (which is guaranteed to be within both convex hulls).} and the algorithm restarts. This description corresponds to a simplified presentation of the algorithm given in~\cite{xie2014maximum}. Previous algorithms (\textit{SwingFilter}~\cite{elmeleegy2009online}, and O'Rourke's one \cite{o1981line}) produce identical results in terms of PLA but may exhibit a longer processing time.

\subsection{Optimal Continuous Approximation} \label{continuous}

Another PLA method is based on continuous approximation where the approximation segments are enforced to form a continuous function. This clearly adds more constraints to the original problem by forcing every consecutive approximation segments to share endpoints (see Figure~\ref{fig:jointknot}) whereas the counterbalance lies in the less information needed to encode the compressed record streams: only two values are needed to store an intersection point, whereas three are required for disjoint knots.

This case is solved optimally by Hakimi and Schmeichel's algorithm~\cite{hakimi1991fitting}. As with the disjoint case, the algorithm efficiently computes and updates two partial convex hulls and two extreme slope lines, but changes the way the approximation is initialized. Instead of starting a new iteration from the break-up point, the algorithm starts the compression from the previous generated line\footnote{Here, this line is carefully chosen to offer the most possibilities.}. Once the next break-up point is found, the method can trace back the origin of the current approximation segment, and only that, one is able to compute reconstructed values for the previous line segment.

\subsection{MixedPLA: a Mixed Approach} \label{mixedPLA}

The last PLA method from the literature tries to combine and balance the benefits offered by both the optimal disjoint and optimal continuous approximation. The essence of this method stands on the pros and cons of the previous PLA methods: continuous PLA offers ``smaller'' compression records but adds an extra constraint to the problem; disjoint PLA is associated with larger compression records but fewer segments are needed than with continuous PLA. Luo \textit{et al} presented in~\cite{luo2015piecewise} a dynamic programming algorithm to find the optimal \textit{size} PLA when the output is a mix of joint and disjoint knots, where each knot type weights as its number of fields. To compute this best global-size PLA, the algorithm uses a mixed approach, combining the two previous optimal procedures within a dynamic programming setting to find the optimal solution. In order to achieve an online solution, the authors have developed an early output approach where fixed segments (belonging without doubt to the optimal mixed-PLA) are output as soon as available.  

\subsection{Linear: Best-fit Approximation} \label{linear}

The last three PLA methods described in this section optimize the PLA stream on one single criterion: number of disjoint knots, number of joint knots, or some weighted function of both numbers. Hence the intrinsic PLA trade-off compression versus loss of accuracy is only encoded in those method as ``compression'' (for one fixed criteria) versus maximum error. Thus individual errors that better characterize the precision loss is lost in the compression. We devise here a simple and efficient novel PLA method aiming at reducing the individual errors by using the best-fit line obtained by a classical simple regression model, while preserving a fixed maximum error. The method follows the greedy procedure: construct the longest possible approximation segment using the best-fit line, and whenever the approximation cannot be pursued, restart from the break-up point.

\begin{figure}[t]\centering
	\begin{tikzpicture}[scale=0.66]
	\node[cross] (0) at (0,2.5) {};
	\draw[|-|] (0,1.5) -- (0,3.5);
	\node[cross] (1) at (1,3.5) {};
	\draw[|-|] (1,2.5) -- (1,4.5);
	\node[cross] (2) at (2,4.16) {};
	\draw[|-|] (2,3.16) -- (2,5.16);
	\node[cross] (3) at (3,4.5) {};
	\draw[|-|] (3,3.5) -- (3,5.5);
	\node[cross] (4) at (4,4.5) {};
	\draw[|-|] (4,3.5) -- (4,5.5);
	\node[cross] (5) at (5,3.5) {};
	\draw[|-|] (5,2.5) -- (5,4.5);
	\node[cross] (6) at (6,1.5) {};
	\draw[|-|] (6,0.5) -- (6,2.5);
	
	\draw[ultra thick] (0,1.5) -- (1,2.5) -- (2,3.16) -- (3,3.5) -- (4,3.5) -- (5,2.5) -- (6,0.5);
	
	
	\draw[thick, dashed, blue, extended line] (0,3.18) -- (6,4.62);
	\draw[thick, dotted, red, extended line] (0,3.74) -- (6,3.14);
	
	\end{tikzpicture}
	\caption{Best-fit lines of a set of points (dashed for $6$ points, dotted for $7$) and associated partial lower convex hull.}\label{fig:linear}
	
\end{figure}
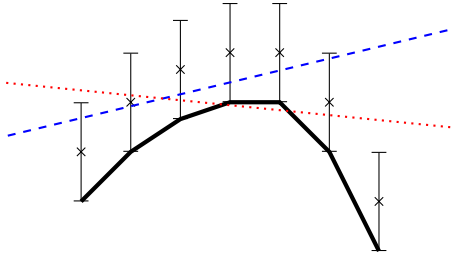

The best-fit line, \textit{i.e.} with slope $a = \hbox{cov}(t,y)/\hbox{var}(t)$ and $y$-intercept $\mu_y - a \mu_t$, is calculated in an online fashion from the covariance $\hbox{cov}(t,y) = (\sum_i t_i y_i)/n - \mu_t \mu_y$, variance $\hbox{var}(t) = (\sum_i t_i^2)/n - \mu_t^2$, averages $\mu_t$ and $\mu_y$ and $n$ the number of compressed points; those quantities are maintained keeping track of the different sums and $n$. Once the best-fit line has been calculated, a second step is to decide whether it is associated with a correct approximation segment for the $n$ points involved, or it violates the error requirement for some of them. The simplest way to check if the computed line is within the error tolerance for all points, is to verify sequentially the approximation error at each currently approximated point. This traversal entails a linear cost to check the validity of the line when $n$ points have been compressed, and may not be appropriate when large portions of the input stream are successfully compressed. To mitigate this issue, we maintain the upper and lower partial convex hulls associated with the current compressed points after each new arrival (as described in \S~\ref{disjoint}). The current best-fit line is then verified by traversing both hulls and checking that the line stays above the lower hull and below the upper one (illustrated by Figure~\ref{fig:linear}). The hulls themselves may have a linear size in $n$, but as already stated, this situation is very unlikely to happen in real traces. 

\section{Streaming PLA}\label{sec:streamingpla}

\subsection{Algorithmic implementations}

In the previous section, we gave an overview of main the PLA methods that we study in this work. Recall the goal of such methods is to produce segment line endpoints in order to approximate the original time series. However, the method itself does not induce how and when should the calculated compression information be output and/or reconstructed. We delegate this task to a separate entity,  called  \emph{streaming PLA compression protocol}, or simply \emph{protocol},  that will be in charge of deciding how and when to output the contents produced by a PLA method. 

The goal of such a protocol is twofold: (1) to produce what we call ``\textbf{compression records}''  from the sequence of PLA records, (2) to provide a method to generate approximation tuples from the stream of compression records. In more detail, the input of a compression protocol is the stream of PLA records produced by some PLA method and the output is a stream of compression records encoding the PLA records in a way to enhance one or several aspects of the compression process (defined in~\S~\ref{subsec:performance_metrics}). Compression records may take different forms depending on the particular protocol; for each such record $r$, $|r|$ represents the relative size of the record compared to a single $y$-value, measured via e.g. its storage space in bytes or the number of fields of $r$. Moreover, with each protocol, a \textit{reconstruction algorithm} is provided, that reads the compression records stream and \textbf{a stream of timestamps}, and produces a stream of reconstructed tuples. As with PLA records, each compression record $r$ is associated with a finite (potentially empty) sequence $\hbox{reconstruct}(r)$ of reconstructed tuples. Each reconstructed record must be within $\varepsilon$ from its original counterpart (reconstructing tuples in sequential order), a property guaranteed by both the correctness of the implied PLA method and the particular protocol used.

As the simplest example of such protocol, we can think of using as compression records directly the PLA records and the reconstruction algorithm given in \S~\ref{sec:preliminaries}. However, we will see in Section~\ref{sec:protocols} that this entails unnecessary extra latency for disjoint knots and that an additional trick is needed for the mixedPLA method to differentiate  the two knot types in the compressed stream.

\subsection{Performance metrics} \label{subsec:performance_metrics}

To evaluate the quality of different PLA methods coupled with various compression protocols, we examine three performance criteria associated with the full PLA compression process, independently of concrete implementation aspects. The three metrics evaluate compression algorithms with respect to three aspects: compression, latency and errors, as argued in the introduction.

Let $M$ be a PLA method (parametrized by the error threshold $\varepsilon$), $P$ a PLA compression protocol and $(t_i,y_i)_{i \geq 0}$ a sequence of $t$-increasing tuples used as input. Furthermore, let $r$ be a compression record, output by $P$ using method $M$ after reading its input stream up to and including timestamp $t_j$. In this situation, we shall denote $\hbox{time}(r) = j$.

Recall, we associate with any compression record $r$ a possibly empty sequence $\hbox{reconstruct}(r)$ of $(t,y)$-tuples that approximates the original input stream. This sequence is computed using a stream of timestamps and the reconstruction procedure associated with $P$. Concatenating the sequences of reconstructed tuples generates the reconstructed stream $(t'_i,y'_i)_{i \geq 0}$. Let finally $\hbox{record}(i)$ be the compression record that produces the $i$-th reconstructed tuple $(t'_i,y'_i)$.

The three metrics considered in this evaluation study are defined for each pair of input tuple $(t_i,y_i)$ and reconstructed tuple $(t'_i,y'_i)$ as follows:

\noindent
\textbf{Compression Ratio}: It corresponds to the ratio $$ \frac{|\hbox{record}(i)|}{|\hbox{reconstruct}(\hbox{record}(i))|} .$$ This ratio between the size of the compression record associated with $i$-th input $\hbox{record}(i)$ and the number of reconstructed tuples generated by $\hbox{record}(i)$ encodes the local quality of compression inspected at the $i$-th input tuple. All reconstructed tuples engendered by the same compression record share the same compression ratio. A ratio greater than~$1$ means that the $i$-th input tuple has been encoded in the compressed stream by more than~$1$ value (see the example in \S~\ref{subsec:implicit}). Any algorithm achieving on average a low compression ratio per point will reduce the number of transmissions needed to propagate a sensor's stream, as well as the storage space needed to store the compressed records stream. This statistic is measured per point to illustrate both sudden bursts in data compression and average data reduction.

\noindent
\textbf{Reconstruction Latency}: It corresponds to the difference $$\hbox{time}(\hbox{record}(i)) - i.$$ This quantity measures the logical latency in terms of the number of tuples $j-i$ that occurred between the input point $(t_j,y_j)$ that triggered the $i$-th reconstructed tuple and the corresponding input tuple $(t_i,y_i)$. Low latency is associated with a fast reconstruction of input tuples (\textit{i.e.} not long after they were received/read). Moreover, an average low latency can be crucial in order to quickly detect and react to anomalies and other important local statistics on the reconstruction side of the compression process. Note a latency of zero is only achieved if after reading $(t_i,y_i)$, a compression record that allows the reconstruction of $(t'_i, y'_i)$ is output; in practice this situation will barely occur, as all PLA methods try to extend the current compression to the next available input tuple, which explains an observed minimum latency of~$1$ input tuple.

\noindent
\textbf{Approximation Error}: It corresponds to the quantity $$|y'_i - y_i|$$ and measures the deviation of the reconstructed tuple from its original counterpart, i.e., it describes how accurate the reconstruction stream is. All approximation errors are within the  parameter $\varepsilon$, however, different PLA methods may have different average. Producing a better compression ratio is naturally associated with pushing the approximation error towards the tolerated threshold. The difficulty of PLA is on finding a good compromise between compression ratio and accuracy loss, and the average approximation error might be a good tool to guide practitioners to decide whether to reduce/increase the error threshold or to change the PLA method used in the streaming-PLA application.

In Section~\ref{sec:evaluation}, we shall analyze these metrics through both the aggregated average and the distribution of the recorded values. Particular attention will be given to peak values, as in our streaming setting, decisions taken are irreversible.
\section{Streaming PLA Protocols}\label{sec:protocols}

In this section, we define precisely how PLA methods can be used to generate a stream of compression records and how those records can be later decoded. We review first the mechanism used in the literature, here named \textsc{Implicit} protocol, and then describe three other streaming protocols focusing on improving different characteristics of the compression process. These three algorithmic implementations will then be associated with different PLA methods and evaluated against our performance metrics, in the evaluation section.

In the following, our compressor scheme is given for $2$-tupled streams; higher dimension streams can be handled by compressing each channel individually. Recall that timestamps are necessary to reconstruct the input stream. When they follow an exact increasing pattern (e.g., for a temperature sensed every 2 minutes), they can be exactly reconstructed at almost no extra cost, by just updating a counter during the reconstruction. We note that in time series used in experimental evaluation in previous works on PLA (eg \cite{elmeleegy2009online, xie2014maximum, luo2015piecewise}), timestamps follow such a regular pattern. Alternatively, the timestamps $(t_i)_{i \geq 0}$ themselves can be compressed as any another channel of the input stream by compressing the stream $(i,t_i)_{i \geq 0}$, with potentially a nil error threshold if uncertainty in reconstructed time is intolerable.

\subsection{Streaming Protocol in Existing Works}\label{subsec:implicit}

In all previous works on PLA except~\cite{luo2015piecewise}, the strategy to output the result of the compression is barely sketched if not eluded. This is easily comprehensible when considering that the optimizing parameter was in each case the number of segments (and perhaps the PLA construction overhead time), and that how those segments were output or stored was of lesser importance. It was therefore left unclear how the segments are represented, while this representation plays a primary role in order to calculate the compression gain.

\begin{figure}[t]\centering
 \begin{tikzpicture}[scale=0.5]
    \node (00) at (-2,2.5) {$\dots$};
    
    \node[cross] (0) at (0,0) {} node[below, xshift=-0.5cm] {$(i,j)$};
    \draw[|-|] (0,1) -- (0,-1);
    
    \node[cross] (1) at (2,2.5) {};
    \draw (1) node[above, xshift=0.75cm] {$(i+1,j')$};
    \draw[|-|] (2,3.5) -- (2,1.5);
    
    \node[cross] (2) at (4,0) {};
    \draw (2) node[below, xshift=-0.75cm] {$(i+2,j)$};
    \draw[|-|] (4,1) -- (4,-1);
    
    \node[cross] (3) at (6,2.5) {};
    \draw[|-|] (6,3.5) -- (6,1.5);
    \draw (3) node[above, xshift=0.75cm] {$(i+3,j')$};
    \node (4) at (8,0) {$\dots$};
    
    \node (5) at (10,2.5) {$\dots$};
    
    \draw[thick, blue] (00) -- (0) -- (1) -- (2) -- (3) -- (4) -- (5);
    \end{tikzpicture}
\caption{Example of worst case compression for the \emph{implicit protocol} (\S\ref{subsec:implicit}), with $j' > j+2\varepsilon$.}\label{fig:badcompress}
\end{figure}
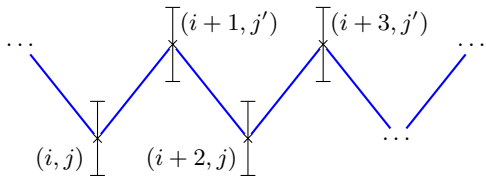

The simplest way, that we call \textit{implicit protocol}, to generate compression records is to output the PLA records themselves, once they have been fully constructed. The reconstruction of values along an approximation segment is then possible as soon as the endpoint of the segment (be it a disjoint or joint knot) is output. It is important to understand the main problem of \textit{inflation} associated with the implicit protocol when the flux of output records has to be streamed. The problem is illustrated by Figure~\ref{fig:badcompress} where an infinite stream of incompressible tuples is received. In this case, if only disjoint knots are allowed, then up to three times more data/transmissions will be needed to encode each single data as a disjoint knot; if joint nodes are used instead, two times more data will be transmitted/stored. In this situation the best achievable compression ratio is~$1$, i.e., no compression and no inflation. One of our protocols, \textsc{TwoStreams} described in \S~\ref{sec:2streams}, ensures data inflation does not occur and can be of interest in streaming situations where the quantity of transmissions cannot be increased by any means.

A second problem concerned latencies associated with a disjoint knot. In PLA methods generating those kind of knots (the optimal disjoint approximation and mixedPLA), the $t$ and $y'$ parameters from a $(t,y',y'')$ disjoint knot are always known before $y''$ (which depends on the compression of the next segment). Hence, the implicit protocol has to always wait one extra segment to reconstruct the previous one. If the algorithm only outputs disjoint knots (except maybe the first one), then this problem is solved by outputting the knot in two parts: first $(t,y')$ then $y''$ when it is computed or equivalently to output triplets of the form $(y'', t', y')$ where $y''$ is the segment $y$-origin and $(t', y')$ is the segment endpoint, and $t'$ is also the $t$-origin of next segment. Nevertheless, mixedPLA still needs an extra adjustment to be able to output both type of knots.

The adjustment is given in~\cite{luo2015piecewise}: compression records for joint knots are identical, \textit{i.e.} of the form $(t,y)$, whereas compression records for disjoint knots use a negative timestamps to represent $t$ and hence are represented by triplets $(-t, y', y'')$. The authors however do not describe how such records can be streamed, and the extra one-segment delay shows up in this case as well. Fortunately, this is again avoidable by outputting disjoint knots $(t,y',y'')$ in two parts: first $(-t,y')$ is generated (where the negative timestamps indicates to the reconstruction algorithm to wait for $y''$), then later on when it has been computed the value $y''$ is generated on its own. What follows depends on the subsequent knot, either a pair $(t',y)$ for a joint knot or a pair $(-t',y)$ for another disjoint knot.

Hence all known PLA methods can be associated with a single common \textsc{Implicit} protocol: joint knots are output when generated, while disjoint knots are yielded in two parts with a negative timestamp. This is the protocol used in our evaluation to measure the performance of the classical PLA methods. 

\subsection{New Streaming Protocols}\label{sec:newprotocols}

In the following three paragraphs, we define other protocols with the aim to perform better in streaming environments. These three protocols would not be able to handle joint knots and thus are compatible only with PLA methods that follow a greedy procedure generating disjoint knots: angle, disjoint and linear. These protocols introduce two main differences with the existing \textit{implicit protocol}:
(1) they feature a mechanism to (at least partially) deactivate PLA compression when it is not working properly, \textit{i.e.} when the error threshold is too small, to avoid data inflation; this is achieved by introducing ``singleton'' tuples, that are not part of any approximation segment. This mechanism is accompanied by a lower bound for PLA compression where segments only encoding a few values are replaced by singletons; (2) they remove (totally or partially) the dependency on representing timestamps (either original ones, or interpolated in case of joint knots) into compression records by using a value $n$ representing the approximation segment length instead; this leads to a potential gain with respect to the compression ratio, as $n$ may need much less space to be represented compared to timestamps (for instance one byte is often enough). With this abstraction, the original (or reconstructed) timestamps are needed to generate the PLA segment stream, but one may notice that the original timestamps are always needed to reconstruct the input stream anyhow. This triggers a need to  bound from above the PLA compression by e.g. $256$ to be able to encode $n$ within single byte. An interesting side-effect is to produce a compression scheme with a maximum bounded latency (here $256$) regardless of how dramatically low the associated compression ratio is. Setting an even lower bound further increases the frequency of output tuples, and features a PLA compression with worst case constant rate feedback even when the compression rate is at maximum.

\subsubsection{Two Streams Protocol}\label{sec:2streams}

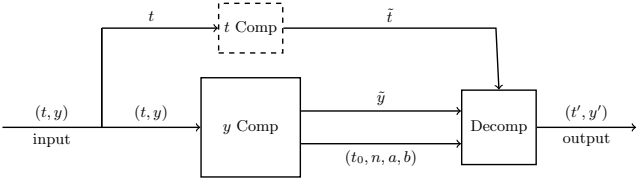
\begin{figure}[t]
	\centering
	\scalebox{0.66}{%
	\centering
	
    \begin{tikzpicture}
	\node[shape=rectangle,black,thick,draw,minimum size=2cm] (c) at (0,0) {$y$ Comp};
	
	\draw[->, thick] (-5,0) -- node[above,xshift=-1cm] {$(t,y)$} node[below,xshift=-1cm] {input} node[above,xshift=1cm] {$(t,y)$} (c);
	
	\node[shape=rectangle,black,thick,draw,minimum size=1cm, dashed] (ct) at (0,2) {$t$ Comp};
	
	\draw[-, thick] (-3,0) -- (-3,2);
	\draw[->, thick] (-3,2) -- (ct);
	\node at (-2,2.25) {$t$};

	\node[shape=rectangle,black,thick,draw,minimum size=1.5cm] (d) at (5,0) {Decomp};
	
	\draw[-, thick] (-3,0) -- (-3,2);
	
	\draw[->, thick] ([yshift=0.33cm]c.east) -- node[above] {$\tilde{y}$} ([yshift=0.33cm]d.west);
	\draw[->, thick] ([yshift=-0.33cm]c.east) -- node[below] {$(t_0,n,a,b)$} ([yshift=-0.33cm]d.west);
	
	\draw[->, thick] (ct.east) -- node[above] {$\tilde{t}$} +(4.28,0) -- (d.north);
	
	\draw[->, thick] (d.east) -- node[above] {$(t',y')$} node[below] {output} ([xshift=2cm]d.east);

	\end{tikzpicture}
    }
	\caption{\textsc{TwoStreams} protocol flowchart.}
	\label{fig:2streams}
\end{figure}

\textsc{TwoStreams} protocol uses two streams of compression records to encode a single input stream of $(t,y)$ tuples (see Figure~\ref{fig:2streams}).

The stream of input tuples $(t,y)$ is half-duplicated such that the $t$ values form a separate stream (which can itself be compressed or not depending on considered applications). In this design, $(t,y)$ tuples are forwarded to a segment line compressor which produces two output streams:
\begin{itemize}[topsep=5pt,itemsep=0pt,parsep=5pt]
	\item[]\textbf{Singletons:} a stream of uncompressed $y$-values;
	\item[]\textbf{Segments:} a stream of line segments represented by quadruplets $(t_0,n,a,b)$ where $t_0$ is the starting time of the segment, $n$ the number of tuples compressed by the line segment, and $(a,b)$ the coefficients of the line approximation to use to reconstruct records.
\end{itemize}

This scheme works in a purely streaming fashion. For each new input tuple, a decision is made whether to output it separately, to compress it with the previous segment or to start a new compression line segment. Each segment is represented by $4$ values and thus a segment line compressing only $3$ input tuples is in this context a waste of space. This is avoided by setting explicitly a minimum compression length of $4$ tuples, redirecting otherwise the $y$-value to the singleton stream.

The reconstruction is performed as follows:
\begin{itemize}[topsep=5pt,itemsep=0pt,parsep=5pt]
	\item if the next timestamp $\tilde{t}$ to process is strictly less than $t_0$, i.e. the  start of the next compressed line segment, then produce the output tuple $(\tilde{t}, \tilde{y})$ by consuming one value $\tilde{y}$ from the \textit{singletons} stream;
	\item on the contrary, if $\tilde{t} \geq t_0$, then reconstruct $n$ tuples $(t_i,y_i)_{1 \leq i \leq n}$ by consuming the next $n$ timestamps $t_1, \dots, t_n$ starting from $\tilde{t}$ and following it, and by setting $y_i = t_i \cdot a + b$.
\end{itemize}

Since segments are only generated for at least $4$ input values, this approach guarantees that the volume of output data is strictly smaller or equal to the input. A drawback of this approach is representing segments by $4$ values and hence consuming more data (here $t_0$, $a$ and $b$ need high precision representations but $n$ can be a byte-long integer). When the input stream is easily compressible, the number of uncompressed values will be fairly low and the cost would be the number of segments times $4$ (counting only number of parameters of the compression records) whereas disjoint knots need only $3$ values.

\subsubsection{Single Stream Protocol}
\begin{figure}[t]
	\centering
    \scalebox{0.75}{%
        \centering
        \begin{tikzpicture}
    	\node[shape=rectangle,black,thick,draw,minimum size=1.5cm] (c) at (0,0) {$y$ Comp};	
    	\draw[->, thick] (-2.5,0) -- node[above] {$(t,y)$} (c);
    	\node[shape=rectangle,black,thick,draw,minimum size=1.5cm] (d) at (5,0) {Decomp};
    	\draw[->, thick] (c.east) -- node[below] {%
    		$\begin{aligned}
    		(n, a, b) &\quad n \geq 3\\
    		(1, y) &\quad
    		\end{aligned}$} (d.west);
    	\draw[->, thick] (d.east) -- node[above] {$(t',y')$} node[below] {output} ([xshift=2cm]d.east);
    	\end{tikzpicture}
    }
	\caption{\textsc{SingleStream} protocol flowchart.}
	\label{fig:singlestream}
\end{figure}
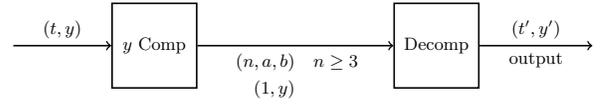

\textsc{SingleStream} protocol is a variant of the first scheme, trying to preserve its advantages (good behavior when PLA compression is low) while optimizing the representation of segments. The single stream protocol is an attempt to deal with the two problems raised by the compression of streaming data. It  represents a compression line segment by only a triplet $(n,a,b)$  where $(a,b)$ are the linear coefficients and $n$ the segment length as in the previous protocol. Singleton values are generated on the same stream preceded by a counter set to~$1$. The lower limit is set to~$3$ input values, whereas the upper limit depends on the number of bytes allocated to represent~$n$ (e.g.~$256$ when~$1$ byte is used).

The reconstruction from compression records is straightforward here: in order to reconstruct the input stream from timestamp $t_i$, check the value of the next counter $n$ on the compression stream. If $n=1$, generate the reconstructed record $(t_i, y)$ where $y$ is the value following the counter; otherwise, when $n \geq 3$ read the two coefficients $a$ and $b$ right after the counter and output the sequence $(t_j,a.t_j+b)_{i \leq j \leq n}$ of $n$ reconstructed tuples consuming along the way the $n-1$ timestamps succeeding $t_i$.

On one hand, this approach may reduce considerably the storage used when the compression of the stream is high since, for each approximated line segment, only $2$ high precision numbers ($a$ and $b$) are required, together with a small counter. This places the protocol with a storage weight per segment strictly between a disjoint ($3$ parameters) and joint knot ($2$ parameters), while removing the extra constraint associated with joint knots (\textit{i.e.} continuous approximation). The worst compression corresponds to Figure~\ref{fig:badcompress}'s scenario where~$1$ extra byte is wasted per input point.

On the other hand, if singleton values are output as soon as they are generated, they can be decompressed right after and do not need to wait for the next approximation segment. This reduces noticeably the latency 
when the PLA compression is poorly effective, \textit{i.e.} when many input tuples are output as singletons rather than within some segment.

\subsubsection{Single Stream Variant Protocol}

With \textsc{SingleStreamV}, a variant of the single stream protocol, approximation line segments are treated in the same fashion, generating triplets of the form $(n,a,b)$, but singleton values are buffered and output in bursts. In more detail, all $m$ singletons $y_1, \dots, y_m$ found between two line segments are combined  into an $m+1$-tuple $(-m, y_1, \dots, y_m)$ where a negative value $-m$ is used to inform the decompression component that the next $m$ values are uncompressed.

Thus the reconstruction from timestamp $t_i$ follows the same scheme as in the single stream protocol for line segments (when the counter $n > 0$) and output the sequence $(t_j,y_j)_{i \leq j \leq -n}$ when $n < 0$ consuming $-n-1$ more timestamps along the way. In this protocol, the lower limit for compression is also set to~$3$ input values, whereas the upper limit is reduced depending on how many ``negative'' values are devoted to represent burst of singletons. If the counter~$n$ uses one single byte, and we do a fair split, the upper limit is set to~$128 \pm 1$ for both segment line and singleton bursts.

In comparison to other protocols, the latency of input tuples here is similar to the two streams protocol when used on low-compression mode. However, having a smaller bound for segments' creation may reduce its compression ratio compare to the same protocol. The worst compression is observed when one singleton is followed by a $3$-values segment where~$2$ counters are needed for~$4$ input values.

\begin{figure}[t]
	\centering
    \scalebox{0.75}{%
        \begin{tikzpicture}
    	\node[shape=rectangle,black,thick,draw,minimum size=1.5cm] (c) at (0,0) {$y$ Comp};	
    	\draw[->, thick] (-2,0) -- node[above] {$(t,y)$} (c);
    	\node[shape=rectangle,black,thick,draw,minimum size=1.5cm] (d) at (6,0) {Decomp};
    	\draw[->, thick] (c.east) -- node[below] {%
    		$\begin{aligned}
    		(n, a, b) &\quad n \geq 3\\
    		(-n, y_1, \dots, y_{n}) &\quad  -n < 0
    		\end{aligned}$} (d.west);
    	\draw[->, thick] (d.east) -- node[above] {$(t',y')$} node[below] {output} ([xshift=1.5cm]d.east);
    	\end{tikzpicture}
    }
    \caption{\textsc{SingleStreamV} protocol flowchart.}
	\label{fig:singlestreamv}
\end{figure}
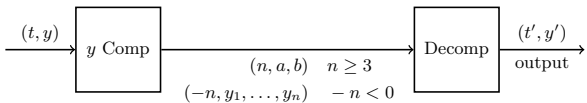

\section{Evaluation}\label{sec:evaluation}

{\em Aim}: We thoroughly evaluate existing PLA techniques against the streaming performance metrics introduced in \S~\ref{subsec:performance_metrics}: compression ratio, approximation latency and individual errors. We evaluate the four classical PLA methods (\textit{SwingFilter}, \textit{SlideFilter}, Optimal continuous and mixedPLA) using the implicit protocol described in \S~\ref{subsec:implicit}, as well as the PLA methods compatible with our protocol framework: Angle (angle-based greedy approximation), Convex Hull (optimal disjoint) and Linear (best-fit line), each associated with our $3$ protocols (\textsc{TwoStreams}, \textsc{SingleStream} and \textsc{SingleStreamV}), in order to find which one offers the best trade-offs for the three streaming performance metrics.
Table \ref{tab:algo} summarizes the methods and protocols, also showing the keys used in the remainder.

{\em Format of results}: We present the results of each couple (metrics/evaluated method) in terms of average value per point and extreme quantiles. Recall that, in our streaming context, discrepancies in the behaviors of the compression algorithm may result in an unbalanceable loss (e.g. sudden high latency or inflation of data).

{\em Comment}: In our streaming scenarios, archetypal data are GPS traces: they are made of timestamped records with multiple channels that draw clear kilometer-long trajectories, and applications usually ask for a target toleration error ranging from a few meters up to about 20 meters independently of the length or range of the trace~\cite{van2015worlds}. Hence, differently from the datasets used in earlier research (sea surface temperature and UCR data set \cite{elmeleegy2009online,xie2014maximum,luo2015piecewise}), our data streams may not follow wave-like patterns at all, and the error threshold is a fixed constant, set beforehand rather than using an error proportional to the $y$-range data stream. 

\begin{table}[t]
    \centering
    \begin{tabular}{c|c|c}
        \textbf{PLA method} & \textbf{Protocol} & \textbf{Key} \\
        \hline
        \multirow{3}{*}{Angle} & \textsc{TwoStreams} & A1 \\
        & \textsc{SingleStream} & A2 \\
        & \textsc{SingleStreamV} & A3 \\
        \hline
        \multirow{3}{*}{Convex Hull} & \textsc{TwoStreams} & C1 \\
        & \textsc{SingleStream} & C2 \\
        & \textsc{SingleStreamV}& C3 \\
        \hline
        \multirow{3}{*}{Linear} & \textsc{TwoStreams} & L1 \\
        & \textsc{SingleStream} & L2 \\
        & \textsc{SingleStreamV} & L3 \\
        \hline
        \textit{SwingFilter} & \multirow{3}{*}{\textsc{Implicit}} & Sw \\
        \textit{SlideFilter} &  & Sl \\
        Optimal Continuous &  & C \\
        MixedPLA &  & M \\
    \end{tabular}
    \caption{Evaluated associations of PLA methods and protocols.}
    \label{tab:algo}
\end{table}

\begin{figure*}[ht!]
\centering
\includegraphics{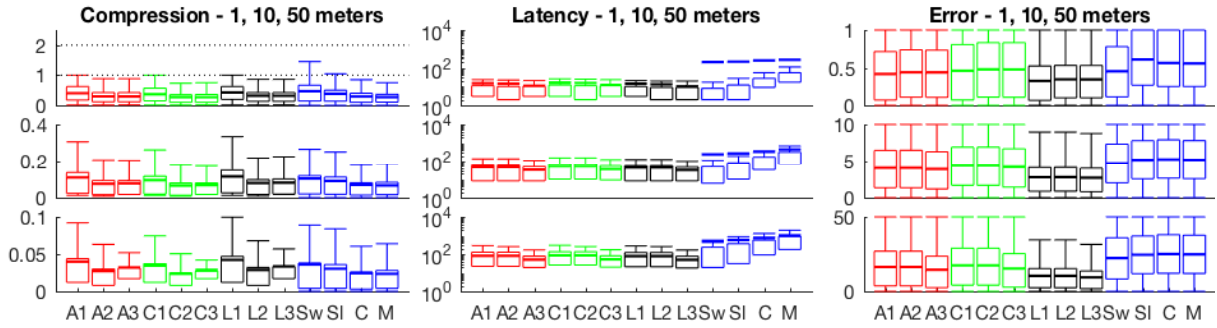}
  \caption{Streaming performance statistics for the GPS dataset and max errors of 1, 10 and 50 meters.}
  \label{fig:gps:comprehensive}
\end{figure*}

\begin{figure*}[ht!]
\centering
\includegraphics{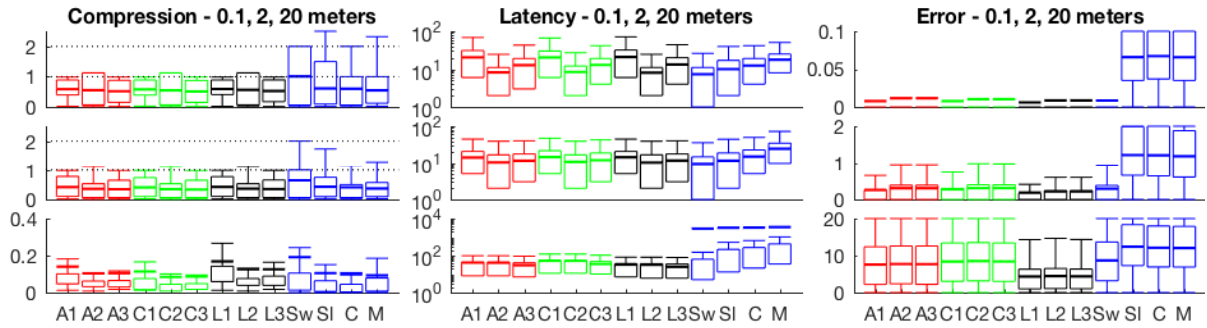}
  \caption{Streaming performance statistics for the LiDAR dataset and max errors of 0.1, 2 and 20 meters.}
  \label{fig:lidar:comprehensive}
\end{figure*}

\begin{figure*}[ht!]
\centering
\includegraphics{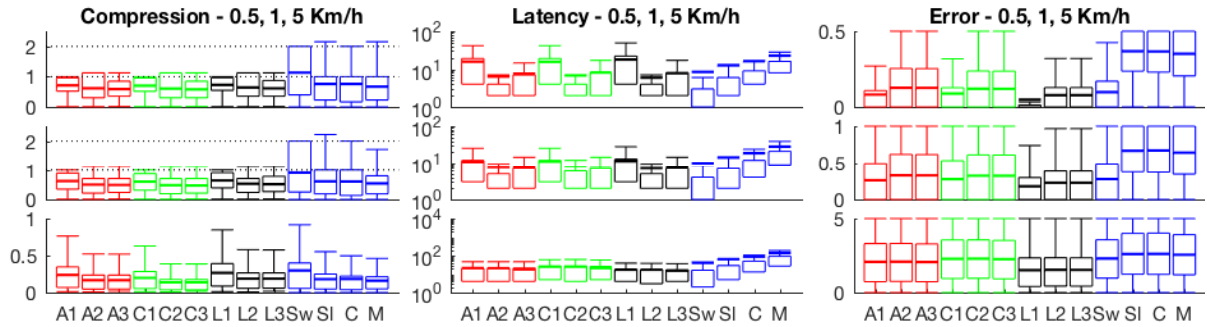}
  \caption{Streaming performance statistics for the URBAN dataset and max errors of 0.1, 1 and 5 meters.}
  \label{fig:here:comprehensive}
\end{figure*}

\begin{figure*}[ht!]
\centering
\includegraphics{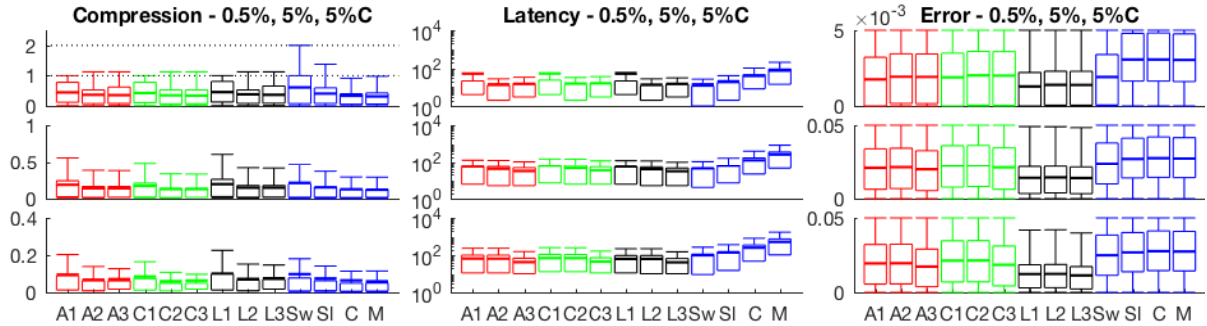}
  \caption{Streaming performance statistics for the UCR dataset and max errors of 0.5\%, 5\% and 5\%C.}
  \label{fig:ucr:comprehensive}
\end{figure*}

\begin{figure*}[ht!]
\centering
\includegraphics{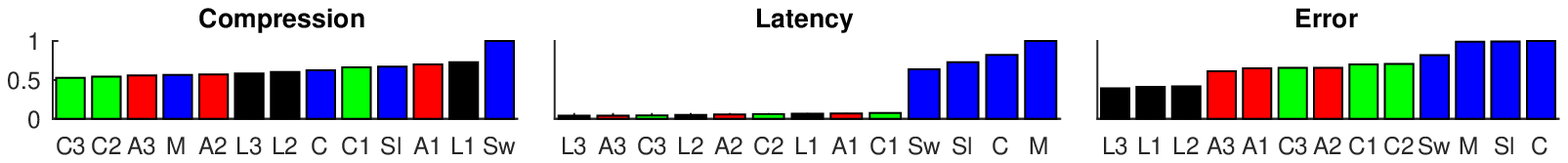}
  \caption{Ranking (from best to worst) of the PLA methods and protocols combinations.}
  \label{fig:summary}
\end{figure*}

\subsection{Datasets}

We use four different datasets to evaluate the presented techniques and protocols:
(a)~The first dataset, named \emph{GPS}, consists of trajectories collected within the scope of the (Microsoft Research Asia) Geolife project by 182 users over a period of approximately three years~\cite{zheng2009mining,zheng2008understanding,zheng2010geolife}.
The dataset we use consists of 17,191 files for each coordinate (X,Y and~Z), containing between 50 and 92,645 lines.
(c)~The second dataset, named \emph{LiDAR}, consists of LiDAR scans from the \textit{Ford Campus} dataset \cite{pandey2011ford}. The LiDAR is a sensor that mounts several lasers on a rotating column. While rotating, these lasers shoot light rays and, based on the time the latter take to reach back the sensor, produce a stream of distance readings. Approximately 1.5 million readings are delivered every second.
The dataset we use consists of 300 files. Each file represents one full rotation of the LiDAR sensor and consists of approximately 150,000 pairs id, distance.
The id value is a logical timestamp that represents the laser that measured the distance and its rotation angle.
(c)~A third dataset named \emph{URBAN}, consists of average vehicle speed measured at different locations in the city of Rio de Janeiro. In total, the data set contains 6,897 files, one for each location. Each file contains approximately 16,200 lines, spanning a time interval of 76 days (readings are taken approximately every 5 minutes).
(d)~For comparing with evaluations in the literature, the fourth dataset we consider is {\em UCR}~\cite{UCRArchive}, using 85 diverse real-world time series from it.

\begin{table*}[ht!]
    \centering
    \begin{tabular}{|>{\centering\arraybackslash}m{1.8cm}|m{3.7cm}|m{10.5cm}|}
        \hline
        \textbf{Statistic} & \textbf{Result} & \textbf{Details} \\
        \hline
        \underline{Compression} & 
        \begin{itemize}[leftmargin=*,topsep=0pt,itemsep=0pt,parsep=0pt]
            \item \textsc{TwoStreams} never inflates data.
            \item \textsc{SingleStream/V} are the best choices for scenario (2).
        \end{itemize}& 
        Under low compression phase, all ``classical methods'' inflate data at different level on a regular basis with a non-negligible portion of the input points being represented by up to slightly more than twice as many bytes. \textsc{TwoStreams} (used in A1, C1 and L1) is the only safe protocol, never showing inflation of data, especially noticeable at low compression phase.
        \textsc{SingleStream/V} provide the best compression ratio only comparable to the mixedPLA method.
        \\
        \hline
        \underline{Latency} & 
        \begin{itemize}[leftmargin=*,topsep=0pt,itemsep=0pt,parsep=0pt]
            \item \textsc{SingleStream/V} exhibit lowest latency.
            \item New protocols are better choices for scenario (1).
        \end{itemize}& 
        Among the three protocols, \textsc{SingleStreamV} exhibits the best latency by using a lower limit for compression segments and burst of incompressible values allowing reconstructed tuples to be output at an earlier stage. The second best protocol is \textsc{SingleStream}, since it avoids the delay involving singletons. All new protocols provides bounded latency (significantly smaller than with classical methods). \\
        \hline
        \underline{Error} & 
        \begin{itemize}[leftmargin=*,topsep=0pt,itemsep=0pt,parsep=0pt]
            \item Linear produce the smallest errors.
            \item New protocols and Linear present better trade-offs compression/error suited for both scenarios.
        \end{itemize} & 
        Our algorithmic implementations, which introduce no error for singleton values, drag the error per point towards $0$. In all observed experiments, most errors due to the approximation are lower than errors created while processing time series using classical PLA methodologies. The PLA method Linear based on a heuristic best-fit line compression presents significant improvement on individual errors while presenting similar delays compared with the other PLA methods, and a compression ratio comparable with the other evaluated PLA methods.\\
        \hline
    \end{tabular}
    \caption{Summary of evaluation results.}
    \label{tab:summary}
\end{table*}

\subsection{Evaluation Setup}

When using one of our three protocols, the compression is bounded to $256$ values per segment and the counter used in segment representation is coded in one byte. All other numbers involved in the input data require at least double precision in order not to break the error threshold just with encoding. Hence, the three fields of disjoint knots, the two fields of joint knots, and each value of the input stream weight $8$ bytes each. The single stream variant protocol sets a limit of $127$ tuples per segment and singleton burst, encoding its counter within one byte.

For each dataset, we have selected three maximum errors in order to compare the different algorithms at a low, medium and high compression levels. For GPS, these are 1, 10 and 50 meters, for LiDAR, 0.1, 2 and 20 meters and for URBAN, 0.5, 1 and 5 km/h.
Since UCR contains diverse time series, we chose the max error as the values representing 0.5\% and 5\% of each file after trimming outliers outside the 5-95\% range, and 5\% considering the entire range (for the latter, we use the label 5\%C in the corresponding graph).

\subsection{Evaluation Results}

The results of the evaluation for each dataset are presented in Figures~\ref{fig:gps:comprehensive} (GPS),  \ref{fig:lidar:comprehensive} (LiDAR), \ref{fig:here:comprehensive} (URBAN) and \ref{fig:ucr:comprehensive} (UCR).
Graphs are composed by three columns and three rows. From left to right, the columns present statistics for compression, latency and error. 
From top to bottom, each row presents the low, medium and high error threshold (also specified in the title of each column). Statistics are shown by a box plot ($25^{th}$ and $75^{th}$ percentiles with whiskers extending up to the extreme points within 1.5 times the difference between the $25^{th}$ and $75^{th}$ percentiles) and a bold line for the average. When the observed compression statistics exceed~1, we plot dotted lines for values~1 and~2 for better readability.
Figure~\ref{fig:summary} presents the ranking of the different PLA methods and protocols (from best to worst) obtained by summing the mean value of each statistics in all the experiments and normalizing them by the highest value.

An overall major comment is that the  {\em most influential factors} on the different metrics are: firstly the protocol used, secondly the PLA method. 
We present in~Table~\ref{tab:summary} the further discussion of the results in a tabular form, to highlight the distilled messages. The table links our results to the two scenarios, (1) sensor transmissions and (2) datacenter storage, that we have identified in \S~\ref{sec:introduction}; \textsc{SingleStream/V} stands for \textsc{SingleStream} and its variant protocols.

\section{Other related work}\label{sec:relatedwork}


Work on PLA dates back to the 1970s~\cite{tomek1974two}, with more recent results in 2014~\cite{xie2014maximum} and 2015~\cite{luo2015piecewise}.  Recall that PLA's main motivation is to reduce representation size of time series by trading data precision. Precision loss can be measured in different ways, the two most common used norms being the euclidean distance $\sqrt{\sum_i (y_i-y'_i)^2}$, or $l_2$-norm, and the maximum error ($l_\infty$-norm). The former one describes an error on the full length of the stream which makes little sense for infinite streams where all decisions have to be made online with solely \textit{apriori} knowledge. Hence, authors have focused more on the latter, which draws a clear \textit{local} condition: every reconstructed data should be within a fixed range $\varepsilon$ from its real counterpart. 


Early works in PLA focused on approximations of 2D-curves (see e.g. \cite{wu1993new} for a brief overview of all techniques) which is a more general problem than our setting. A notable early algorithm  \cite{tomek1974two} already provides in 1974 a constant-time solution to compute heuristically a PLA representation of a single variable function (exactly the problem studied here).

The continuous PLA problem is solved optimally by Hakimi and Schmeichel~\cite{hakimi1991fitting} in 1991 with a variation of the original algorithm~\cite{imai1986optimal} from 1986. In 2009, Elmeleegy \textit{et al} \cite{elmeleegy2009online} introduce the \textit{SlideFilter} and \textit{SwingFilter} algorithms. As noted in \cite{xie2014maximum}, \textit{SlideFilter} happens to have been already published in 1981 by O'Rourke \cite{o1981line} while the angle-based algorithm \textit{SwingFilter} appears explicitly as early as 1983 in \cite{gritzali1983fast}. Note also the widely used SWAB algorithm~\cite{keogh2001online} that focuses on heuristically optimizing the $l_2$-error norm of the PLA.

In 2014 \cite{xie2014maximum}, the authors describe a novel optimal disjoint PLA method. An experimental study is conducted to show that reducing the convex hulls size by using the bounding slope lines is a good performance trade-off leading to diminishing processing time. The work has since been extended to cover also the continuous case in \cite{zhao2016segmenting}. Zhao \textit{et al} show that after running the convexhull method (tuned to produce continuous segmentation), we can check if two consecutive disconnected segments (using their respective hulls) can be merged into two consecutive joint segments, a sufficient strategy to achieve better but not optimal segmentation. Clearly the authors were unaware that the optimal case has indeed previously been solved \cite{hakimi1991fitting}.

The closest work sharing some of our preoccupations appears in 2015 \cite{luo2015piecewise}. Luo \textit{et al} remark that minimizing the total number of segments might not effectively minimize the needed space to store the PLA representation. They hence introduce the nomenclature repeated in this article: joint knots $\langle x,y \rangle$ and disjoint knots $\langle x,y_1,y_2 \rangle$; a sign trick (storing $-x$ instead of $x$) is used not to have to store an additional bit while streaming output tuples. The paper presents a dynamic programming scan-along procedure to optimize the ``size'' of the approximation, where for each approximation segment a joint knot is counted as $2$ whereas a disjoint knot as $3$. For the first time, the space required by the PLA is not measured in number of segments. Also, aware that their dynamic programming solution slows down the output rate, the authors measured the maximum observed delay in terms of number of segments and claim outputs in the worst case in experiments less than to $2$ to $4$ segments after each point's processing. Also, the $\ell_2$-error is measured to quantify the loss in precision with different PLA methods. We have extended their reasoning one step further by actually measuring the number of bytes needed by each parameter forming compression records, and by measuring delays and errors at a more meaningful point level.

Several of these previous works (\cite{xie2014maximum,luo2015piecewise,zhao2016segmenting}) use the UCR time series data \cite{UCRArchive} as a benchmark. As already argued, due to its heterogeneity we do not think this dataset is a meaningful example in our streaming setting but we have included it as well in our evaluation for seek of comparison.

Finally, let us note that some recent related work studied the compression of time-series, e.g. \cite{geurts2001pattern, keogh2001locally, lin2003symbolic}, using different schemes including, among others, piecewise linear and piecewise constant approximation in order to solve different problems like classification of time series, and time series data mining. Since in constant piecewise approximation the average of the stream is used to approximate portions of it, no bounding error is given for each individual input value but instead the error along a segment is captured by the euclidean distance. As shown in these works, piecewise constant approximation can be the best tool to use in other applications than ours. For further references on variations of the PLA problem and its applications in time series mining, see the complete survey from~\cite{Ratanamahatana2010}.
\section{Discussion and Future Work}\label{sec:conclusions}

In this experimental survey about usage of PLA  in realistic streaming situations, we have identified two scenarios where PLA can be used to reduce the data volume: either at one edge of the network by reducing the number of transmissions of a sensor, either at the other edge by reducing the storage space required to record some received streams.

To measure the quality of compression achieved in streaming environments by state-of-the-art PLA techniques, we formalized three performance metrics taking into account data reduction performance, but also latency introduced along the compression as well as the precision lost in the process.

Along those metrics, we have presented a complete streaming framework to be able to evaluate PLA methods, and ``compression protocols'' that may be used with different methods. 
This framework can ease further research with common criteria for the compression's performance.

Our main contribution is an evaluation of 13 combinations of PLA methods and streaming protocols on four real-world datasets of timestamped numerical streams. The evaluated procedures include existing PLA methods and different streaming techniques that we have introduced to improve their performance. Results show that for comparable compression ratios, our techniques reduce significantly both the delays on the reconstructed stream and the individual errors resulting from PLA compression. 

This work highlights some of the challenges raised by using PLA within modern streaming applications. The PLA method in general relies on fixing a maximum error that can be tolerated and then tries to optimize what level of compression can be achieved within the tolerated error. However, streamed data may suddenly change in a radical way due to many factors, and setting a too tight threshold often results in inflation of the data volume, nullifying PLA's purpose. Nevertheless, we have demonstrated here that small variations in the way PLA records are computed and generated allow to improve performance both on a low and high compression mode. A promising way to extend this work is to perform such adjustment automatically, \textit{i.e.} to exhibit adaptive methods capable of changing the way PLA is yielded to preserve the best possible overall performance (a high compression with small reconstruction delays).

Another aspect not addressed by this evaluation concerns the computational overhead associated with PLA compression. All introduced methods are sufficiently simple to result in a small processing overhead on the compression side. It would be interesting, however, to confirm this claim for realistic deployment scenarios.

\balance


\bibliographystyle{abbrv}
\bibliography{bibtex}  



%

\end{document}